\documentclass[aps,prl,10pt,twocolumn,showpacs,preprintnumbers,amsmath,amssymb,superscriptaddress]{revtex4-1}
\usepackage[english]{babel}
\usepackage[dvips]{graphicx}
\usepackage{dcolumn}
\usepackage{bm}

\usepackage{color}

\begin{document}
\title{A table-top laser-based source of femtosecond, collimated,\\ ultra-relativistic positron beams}
\author{G. Sarri}
\affiliation{School of Mathematics and Physics, The Queen's University of Belfast, BT7 1NN, Belfast, UK}
\author{W. Schumaker}
\affiliation{Center for Ultrafast Optical Science, University of Michigan, Ann Arbor, Michigan 48109-2099, USA}
\author{A. Di Piazza}
\affiliation{Max-Planck-Institut f\"{u}r Kernphysik, Saupfercheckweg 1, 69117 Heidelberg, Germany}
\author{M. Vargas}
\affiliation{Center for Ultrafast Optical Science, University of Michigan, Ann Arbor, Michigan 48109-2099, USA}
\author{B. Dromey}
\affiliation{School of Mathematics and Physics, The Queen's University of Belfast, BT7 1NN, Belfast, UK}
\author{M. E. Dieckmann}
\affiliation{School of Mathematics and Physics, The Queen's University of Belfast, BT7 1NN, Belfast, UK}
\author{V. Chvykov}
\affiliation{Center for Ultrafast Optical Science, University of Michigan, Ann Arbor, Michigan 48109-2099, USA}
\author{A. Maksimchuk}
\affiliation{Center for Ultrafast Optical Science, University of Michigan, Ann Arbor, Michigan 48109-2099, USA}
\author{V. Yanovsky}
\affiliation{Center for Ultrafast Optical Science, University of Michigan, Ann Arbor, Michigan 48109-2099, USA}
\author{Z. H. He}
\affiliation{Center for Ultrafast Optical Science, University of Michigan, Ann Arbor, Michigan 48109-2099, USA}
\author{B. X. Hou}
\affiliation{Center for Ultrafast Optical Science, University of Michigan, Ann Arbor, Michigan 48109-2099, USA}
\author{J. A. Nees}
\affiliation{Center for Ultrafast Optical Science, University of Michigan, Ann Arbor, Michigan 48109-2099, USA}
\author{A. G. R. Thomas}
\affiliation{Center for Ultrafast Optical Science, University of Michigan, Ann Arbor, Michigan 48109-2099, USA}
\author{C. H. Keitel}
\affiliation{Max-Planck-Institut f\"{u}r Kernphysik, Saupfercheckweg 1, 69117 Heidelberg, Germany}
\author{M. Zepf}
\affiliation{School of Mathematics and Physics, The Queen's University of Belfast, BT7 1NN, Belfast, UK}
\affiliation{Helmholtz Institute Jena, Fr\"{o}belstieg 3, 07743 Jena, Germany}
\author{K. Krushelnick}
\affiliation{Center for Ultrafast Optical Science, University of Michigan, Ann Arbor, Michigan 48109-2099, USA}
\date{\today}
\begin{abstract}
The generation of ultra-relativistic positron beams with short duration ($\tau_{e^+} \leq 30$ fs), small divergence ($\theta_{e^+} \simeq 3$ mrad), and high density ($n_{e^+} \simeq 10^{14} - 10^{15}$ cm$^{-3}$) from a fully optical setup is reported. The detected positron beam propagates with a high-density electron beam and $\gamma$-rays of similar spectral shape and peak energy, thus closely resembling the structure of an astrophysical leptonic jet. It is envisaged that this experimental evidence, besides the intrinsic relevance to laser-driven particle acceleration, may open the pathway for the small-scale study of astrophysical leptonic jets in the laboratory.
\end{abstract}
\pacs{}
\maketitle

Creating and characterizing high-density beams of relativistic positrons in the laboratory is of paramount importance in experimental physics, due to their direct application to a wide range of physical subjects, including nuclear physics, particle physics, and laboratory astrophysics. Arguably, the most practical way to generate them is to exploit the electromagnetic cascade initiated by the  propagation of an ultra-relativistic electron beam through a high-$Z$ solid. This process is exploited to generate low-energy positrons in injector systems for conventional accelerators such as the Electron-Positron Collider (LEP)  \cite{LEP}. In this case, an ultra-relativistic electron beam ($E_{e^-}\approx 200$ MeV) was pre-accelerated by a LINAC and then directed onto a tungsten target. The resulting positron population, after due accumulation in a storage ring,  was further accelerated by a conventional, large-scale ($R\approx 27$ km), synchrotron accelerator up to a peak energy of 209 GeV. The large cost and size of these machines have motivated the study of alternative particle accelerator schemes. A particularly compact and promising system is represented by plasma devices which can support much higher accelerating fields (of the order of 100s of GV/m, compared to MV/m in solid-state accelerators) and thus significantly shorten the overall size of the accelerator. Laser-driven generation of electron beams with energies per particle reaching  \cite{Kneip,Leemans,Hafz,Lundh}, and exceeding \cite{Wang}, 1 GeV have been experimentally demonstrated and the production of electron beams with energies approaching 100 GeV is envisaged for the next generation of high-power lasers (1 - 10 PW) \cite{Tzoufras}. Hybrid schemes have also been proposed and successfully tested in first proof-of-principle experiments \cite{Blumenfeld,Muggli}. On the other hand, laser-driven low energy positrons ($E_{e^+}\approx 1 - 5$ MeV) have been first experimentally obtained by C. Gahn and coworkers \cite{Gahn} and recently generated during the interaction of a picosecond, kiloJoule class laser with thick gold targets \cite{Chen1, Chen2,Cristoph, Tonino}. Despite the intrinsic interest of these results, the low energy and broad divergence reported ($E_{e^+}\leq 20$ MeV and $\theta_{e^+} \geq 350$ mrad , respectively) still represent clear limitations for future use in hybrid machines.

The possibility of generating high density and high energy electron-positron beams is of central importance also for astrophysics, due to their similarity to jets of long gamma-ray bursts (GRBs), which are ejected as a consequence of the collapse of super-massive stars \cite{Woosley}. These structures, despite extensive numerical studies \cite{Silva}, still present enigmatic features which are virtually impossible to address by simply relying on direct observations. A possible solution might be represented by reproducing small scale electron-positron jets (required bulk flow Lorentz factor of the order of 100 - 1000) in the laboratory. Although GRB jets may well have a weak large scale magnetic field \cite{Wiersema}, the external shock is exclusively mediated by self-generated micro-scale magnetic fields. A purely electronic jet would present toroidal magnetic fields whose strength and structure would be comparable to the microscale fields that develop in response to the filamentation instability \cite{Medvedev} and modify the shock physics. The presence of the highly mobile positrons would reduce the overall magnetisation of the jet (alongside with the amplitude of the electrostatic fields driven by a charge separation \cite{Dieckmann}), simplifying the interpretation of the experimental data and their comparison with the astrophysical scenario.

\begin{figure*}[!t]
\begin{center}
\includegraphics[width=2\columnwidth]{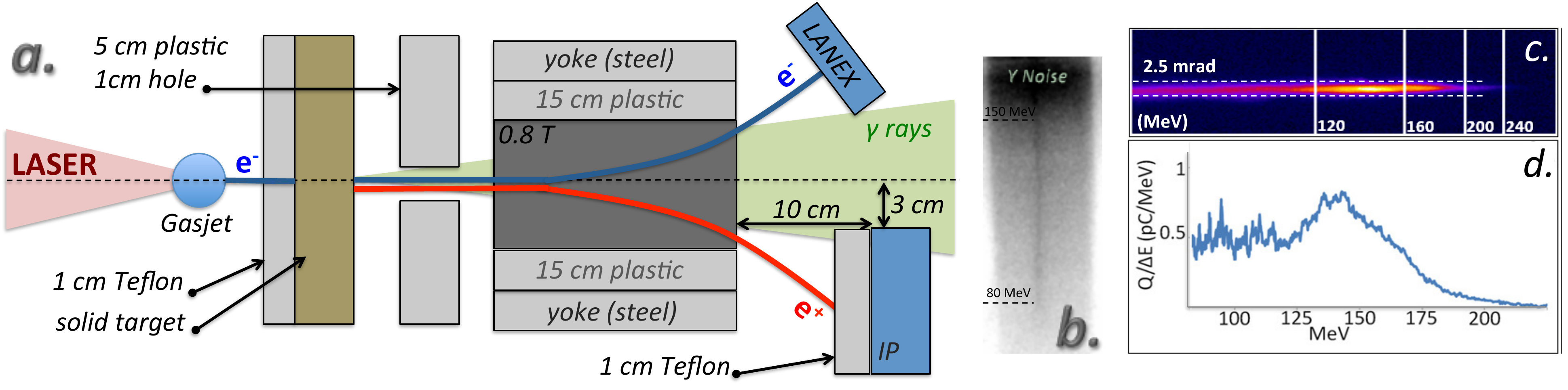}
\caption{\textbf{a.}  Top-view of the experimental setup. Plastic and Teflon shielding was inserted to reduce the noise due to low energy divergent particles and x-rays.  \textbf{b.} Typical positron signal as recorded by the Image Plate. The region labelled with $\gamma$ noise is predominantly exposed by the $\gamma$-rays escaping the solid target. \textbf{c.} Typical signal of the electron beam as recorded on the LANEX screen, without a solid target and  \textbf{d.} extracted spectrum.}
\label{setup}
\end{center}
\end{figure*}
Here we experimentally demonstrate the possibility of producing collimated and high-density ultrarelativistic positron beams in a fully laser-driven configuration. Their low divergence and short duration (comparable to those of the incoming laser-driven electron beam) suggest the possibility of applying this generation scheme to current laser facilities towards the construction of a fully-optical generator of high-quality, ultra-relativistic positron beams. Furthermore,the measured high positron Lorentz factors ($\gamma_{e^+}\simeq 200 - 400$, compared to $\gamma_{e^+}\leq 10$ in \cite{Gahn,Chen1,Chen2}) of these beams are finally comparable to those of astrophysical leptonic jets. This, in conjunction with the low divergence, the inferred electron/positron balance, and co-propagation with intense gamma-rays, finally open up a realistic possibility of studying the dynamics of such jets in the laboratory.

The experiment (shown schematically in Fig. \ref{setup}.a) was carried out using the HERCULES laser system at the Center for Ultrafast Optical Science (CUOS) in the University of Michigan \cite{HERCULES}, which delivered a laser beam with a central wavelength $\lambda_L=0.8$ $\mu$m, energy $E_L=0.8$ J and duration $\tau_L=30$ fs. This laser beam was focussed, using an $f/20$ off-axis parabola, onto the edge of a 3 mm wide supersonic He gas-jet, doped with 2.5\% of N$_2$, with a backing pressure of 5.5 bar. Once fully ionised, this corresponds to an electron density of $9\times10^{18}$ cm$^{-3}$. The focal spot size was measured to have a radius of 23 $\mu$m which contained 50\% of the laser energy (peak intensity of $I_L\approx6\times10^{18}$ W/cm$^{2}$). Laser power and gas-jet pressure were chosen in order to stay slightly above the threshold for ionisation injection \cite{ionization}. This interaction delivered a reproducible electron beam with a divergence of approximately 2.5 mrad (see Figs. \ref{setup}.d and \ref{setup}.e). Its spectrum was measured, prior to any shot with a high-$Z$ solid target, by a magnetic spectrometer consisting of  a 0.8 T, 15 cm long pair of magnets and a LANEX screen. The arrangement of the spectrometer did not allow us to resolve electron energies below 80 MeV. Typical spectra, obtained using the calibration curves reported in \cite{Glinec}, indicated the charge carried by electrons with energy exceeding 80 MeV to be of the order of 50 pC ($3\times10^8$ electrons). Electron bunches obtained in similar conditions have been shown to have a length comparable to a plasma wavelength ($\lambda_{pe}= 2\pi c/\omega_{pe} \approx 10$ $\mu$m) implying a typical temporal duration comparable to that of the laser pulse \cite{Mangles}.The laser-accelerated electron beam interacted with mm-size high-$Z$ solid targets of different materials (Cu, Sn, Ta, Pb) and thicknesses (from 1.4 to 6.4 mm). The same magnetic spectrometer was used to separate the electrons from the positrons which were then recorded onto an Image Plate (IP). Due to the small difference in positron and electron stopping power (below 2\% \cite{Rohrlich}), the signal recorded was absolutely calibrated by using the calibration curves reported in \cite{Tanaka}. Plastic shielding was inserted to reduce the noise on the IP induced by both the low-energy electrons and gamma-rays generated, at wide angles, during the laser-gas and electron-solid target interactions (see Fig. \ref{setup}.a).

\begin{figure}[!t]
\begin{center}
\includegraphics[width=1\columnwidth]{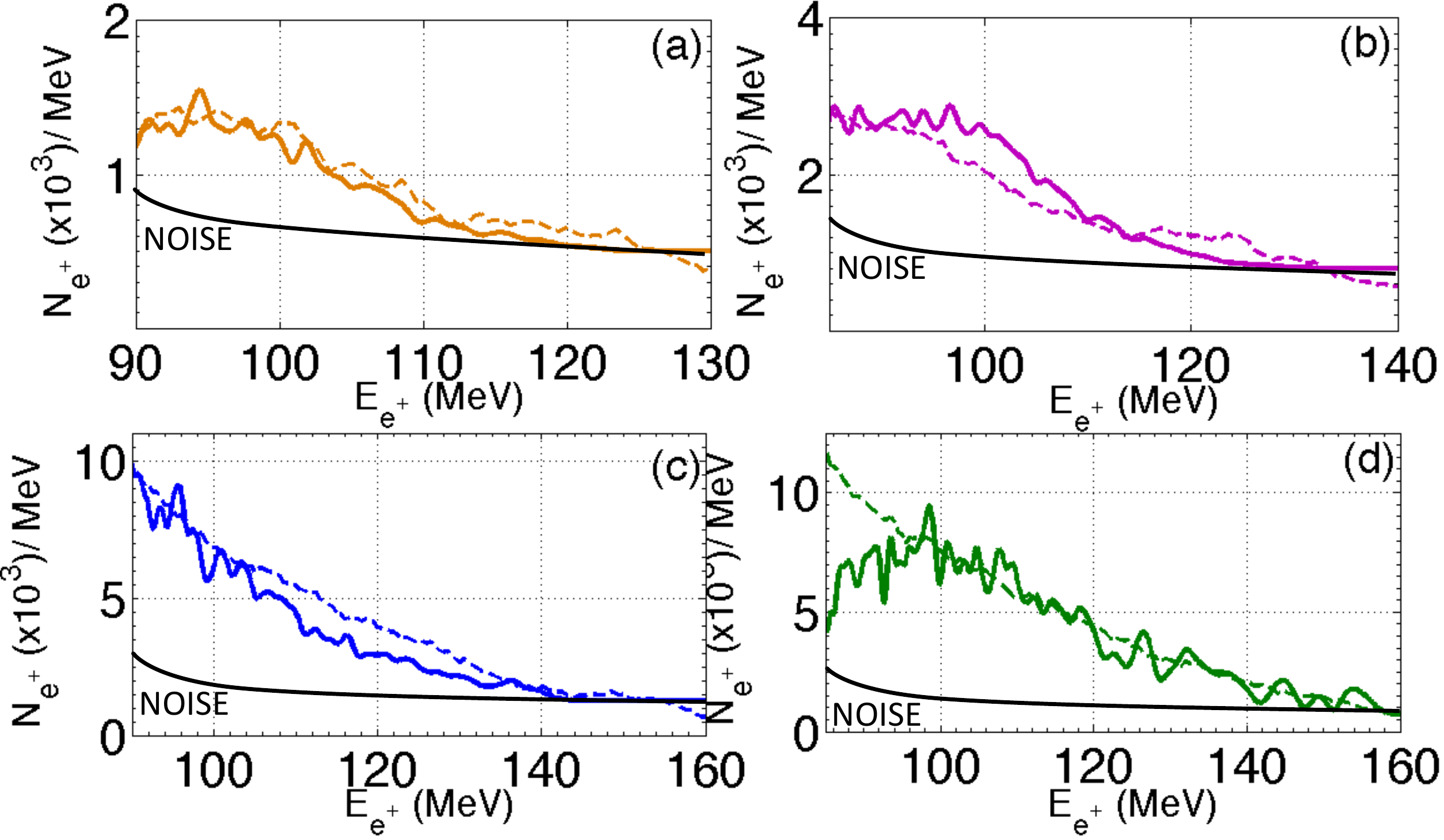}
\caption{ Experimental (solid lines) and simulated (dashed lines) positron spectra for \textbf{(a)} 5.3 mm of Cu, \textbf{(b)} 6.4 mm of Sn, \textbf{(c)} 2.8 mm of Ta and \textbf{(d)} 4.1 mm of Pb.} \label{spectra}
\end{center}
\end{figure}

\begin{figure}[!t]
\begin{center}
\includegraphics[width=1\columnwidth]{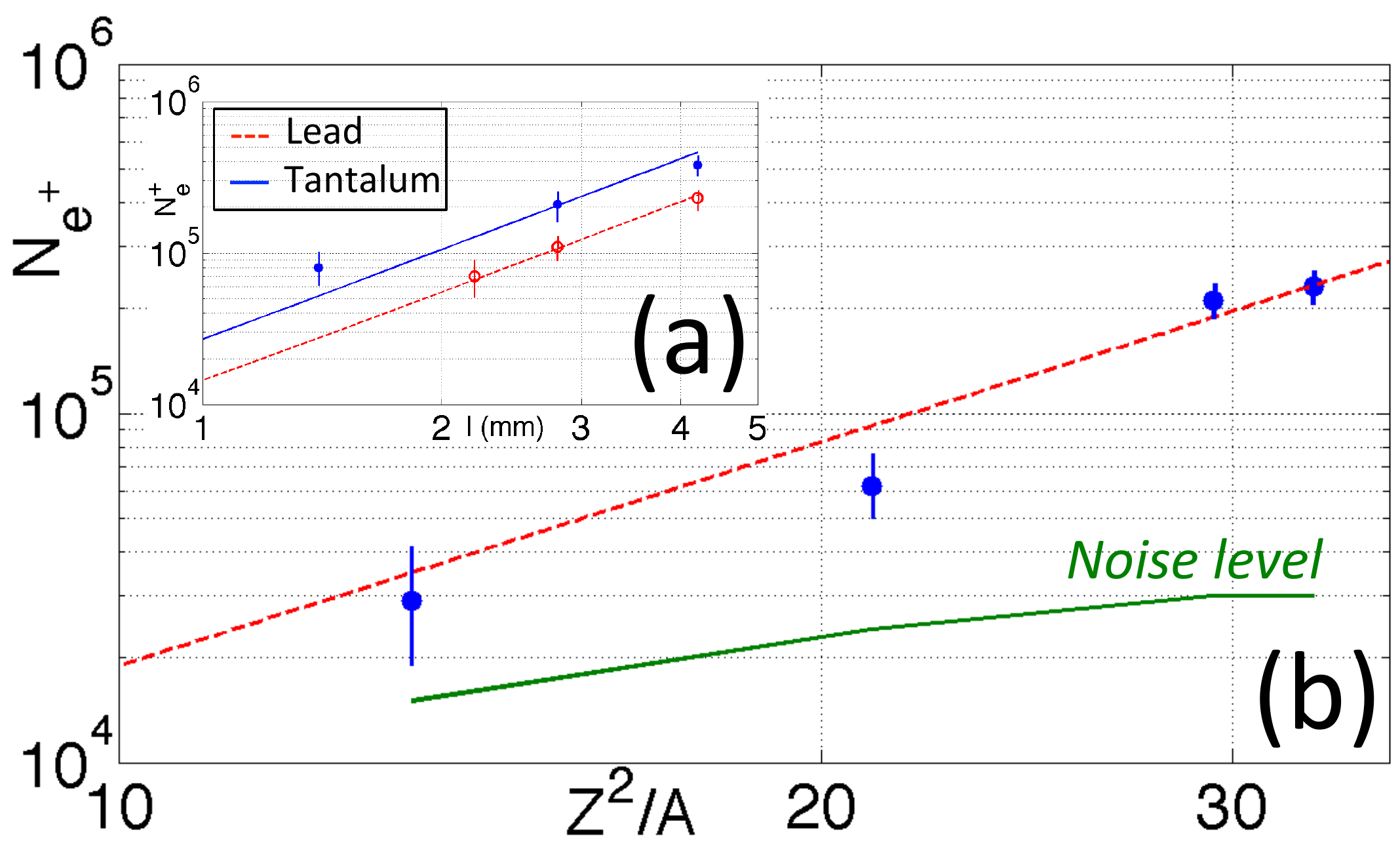}
\caption{\textbf{(a)} Measured positron yield, in the energy range $90<E_{e^+}$(MeV)$<120$, for Ta (blue full circles) and Pb (red empty circles) for different target thicknesses. Lines give the best quadratic fits.   \textbf{(b)} Measured positron yield, in the energy range $90<E_{e^+}$(MeV)$<120$, for different materials but constant areal density as a function of $Z^2/A$. The dashed line represents the best quadratic fit.} \label{general}
\end{center}
\end{figure}

In these experimental conditions, the positrons inside the high-Z target can be generated via either direct electroproduction (trident process), in which pair production is mediated by a virtual photon in the electron field \cite{Baier}, or via a two-step ``cascade'' process where the electron first emits a real photon (bremsstrahlung) \cite{Koch} which then produces an electron/positron pair via the Bethe-Heitler process \cite{Heitler}. Higher-order multi-step cascade processes may also significantly contribute to pair production, depending on the ratio between the thickness of the target and the radiation length of the material \cite{Landau}. Keeping the parameters of the electron beam constant, the positron yield $N_{e^+}$ is expected to scale as: $N_{e^+}\propto (Z^2n d)^j$,  where $n$ is the number of atoms per unit volume in the material, $d$ is the thickness of the solid target, and $j=1$ for the trident process and $j=2$ for the two-step cascade process (we neglect here Coulomb corrections, which depend on $Z\alpha$, with $\alpha\approx 1/137$ being the fine-structure constant). Neglecting the difference between the proton and the neutron mass, the mass density of the solid target is $\rho\approx Am_pn$, with $A$ and $m_p$ being the atomic number and the proton mass, respectively. If we maintain the areal mass density ($\sigma=\rho d$) constant, we can then express the scaling as $N_p\propto (Z^2/A)^j $. We have thus performed a series of shots for different materials (Cu, Sn, Ta, Pb) adjusting the target thickness so that the areal mass density was kept constant for each material ($\sigma\approx 4.7$ g/cm$^2$, see first four rows in Table \ref{tab0}). All the measured positron spectra, inferred from the signal recorded on the IP (which was absent if no solid target was inserted in the electron beam path, see Supplementary Material), showed a monotonically decreasing profile with approximately $10^3$ positrons/MeV (solid lines in Fig. \ref{spectra}). 
In all cases, numerical simulations performed with the nuclear physics Monte-Carlo code FLUKA, which accounts for electromagnetic cascades during the passage of an electron beam through a solid target \cite{FLUKA}, are able to reproduce the experimental data well (dashed lines in Fig. \ref{spectra}). As theoretically predicted, the positron yield increases for materials with higher atomic number. This trend is quantitatively confirmed by integrating the experimental spectra in the range $90<E_{e^+}$(MeV)$<120$ (see Table \ref{tab0} and Fig. \ref{general}). Within this energy range, a maximum positron number of $(2.30\pm0.28)\times10^5$ is obtained for the material with the highest $Z$ (Pb). Keeping $j$ as a free parameter, we obtain these data to be best fitted, as a function of $Z^2/A$, if $j=2.1\pm0.1$ confirming the dominance of the cascade process with respect to the trident one (see Fig. \ref{general}.b). The positron yield over the entire positron spectrum, as extracted from matching FLUKA simulations ($N_T$ in Table \ref{tab0}), is seen to follow a similar trend. A further indication as to what process dominates is given by the dependence of the positron yield upon the target thickness ($N_{e^+}\propto d$ for the trident and $N_{e^+}\propto d^2$ for the two-step process). A series of shots was thus taken varying the thickness of the solid target $d$ for Ta and Pb (second four rows in Table \ref{tab0} and Fig. \ref{general}.a). As expected, the positron yield is seen to scale with $d^2$ in both cases. In order to support this statement theoretically we compare, for each material, the radiation length $L_{\text{rad}}$ with the range of target thicknesses $d$ used in the experiment. The contribution of two-step process is expected to exceed the one of the trident process if $d/L_{\text{rad}}\gtrsim 10^{-2}$ \cite{Baier}. For an order-of magnitude estimate of $L_{\text{rad}}$, we can assume here to be in the total-screening regime. For an electron with energy $\varepsilon$ emitting a photon with energy $\hbar\omega$, there is total screening if the parameter $S\equiv \alpha Z^{1/3}\varepsilon(\varepsilon-\hbar\omega)/\left(\hbar\omega mc^2\right)$ is much larger than unity, where the Thomas-Fermi model of the atom has been assumed \cite{Landau}. In order to qualitatively explain the reported experimental results, we can roughly estimate $\varepsilon\sim\hbar\omega\sim 100\;\text{MeV}$. Thus, it is $S\gtrsim 4$ in the worst case of Cu, which is sufficiently large for the present estimate. In this regime, and by including Coulomb corrections, the radiation length is approximately given by \cite{Landau}
\begin{equation}
\label{L_rad}
L_{\text{rad}}=\frac{1}{4\alpha(Z\alpha)^2n\lambda^2_C L_0},
\end{equation}
where $n$ is the number of atoms per unit volume, $\lambda_C=\hbar/mc=3.9\times 10^{-11}\;\text{cm}$ is the Compton wavelength, and $L_0=\log(183Z^{-1/3})-f(Z\alpha)$, with $f(x)=\sum_{\ell=1}^{\infty}x^2/\ell(\ell^2+x^2)$. Eq. (\ref{L_rad}) provides the following values for the materials employed in the experiment: $L_{\text{rad}}(\text{Cu})=15\;\text{mm}$, $L_{\text{rad}}(\text{Sn})=12\;\text{mm}$, $L_{\text{rad}}(\text{Ta})=4.1\;\text{mm}$ and $L_{\text{rad}}(\text{Pb})=5.6\;\text{mm}$. As seen by comparing these values with those in Table I, the material thicknesses are always such that the inequality $d/L_{\text{rad}}\gtrsim 10^{-2}$ is fulfilled, in agreement with the experimental indication of the predominance of a two-step process for the electromagnetic cascade. Moreover, in all the considered cases, except one where $d=4.2\;\text{mm}$ for Ta, it is $d<L_{\text{rad}}$, which implies that the contribution of higher-order cascade processes can generally be neglected for an order-of-magnitude estimate. This is also corroborated by the observed angular divergence of the positron beams of the order of $1/\gamma$, with $\gamma\approx 300$ being the relativistic factor of the incoming electron beam (a larger number of cascade steps implies in general a broader angular distribution of the final particles).
\begin{table}[!b]
\begin{center}
\begin{tabular}{|c|c|c|c|c|c|}
\hline
Mat. & d(mm) & $\theta_{e^+}$(mrad) & $N_{\mbox{exp}}$ $\times10^5$ & $N_{\mbox{sim}}$ $\times10^5$ & $N_{T}\times10^5$\\
\hline
Cu & 5.3 & $2.3\pm0.2$ & $0.3\pm0.1$ & 0.3 & $31$\\
Sn & 6.4 & $2.7\pm0.3$ & $0.6\pm0.1$ & 0.6 & $63$\\
Ta & 2.8 & $2.7\pm0.3$ & $2.1\pm0.3$ & 2.1 & $190$\\
Pb & 4.2 & $3.5\pm0.4$ & $2.3\pm0.3$ & 2.3 & $240$\\
\hline
Ta & 1.4 & $2.3\pm0.2$ & $0.8\pm0.2$ & 0.8 &$78$\\
Ta & 4.2 & $2.7\pm0.3$ & $3.8\pm0.3$ & 3.9& $350$\\
Pb & 2.2 & $3.0\pm0.3$ & $0.7\pm0.2$ & 0.7& $60$\\
Pb & 2.8 & $3.3\pm0.3$ & $1.1\pm0.3$ & 1.1 & $122$\\
\hline \hline
\end{tabular}
\end{center}
\caption{The first four rows illustrate the results from targets with same areal density. The positron yield $N_{\mbox{exp}}$ and $N_{\mbox{sim}}$ refer to the energy window $90 < E_{e^+}$(MeV)$ < 120$ as obtained from the experiment and FLUKA simulation, respectively. $N_{T}$ refers instead to the total yield of positrons  with $E_{e^+}>1$MeV, as extracted from matching numerical simulations.}\label{tab0}
\end{table}

Due to the divergence of the positron beam, its maximum density is located at the close vicinity of the rear side of the solid target.  Here, the positron beam has a transverse diameter of the order of $150$ $\mu$m and, by assuming that the positron beam will retain the temporal duration of the initial electron beam ($\tau_{e^-} \leq 30$ fs, \cite{Mangles}), a longitudinal length of the order of $c\tau_{e^-} \leq 9$ $\mu$m. In the case of maximum yield (4.2 mm Ta, see Table \ref{tab0}) the density of positrons with an energy between 90 and 120 MeV is of the order of $2.3\times10^{12}$ cm$^{-3}$. FLUKA simulations indicate that this energy window contains approximately 1\% of the total positron yield. For 4.2 mm of Ta, this means that the total amount of positrons with energy $E_{e^+} > 1$ MeV will be of the order of $3.5 \times 10^7$, indicating an overall positron density of about $2\times10^{14} $ cm$^{-3}$. The overall positron beam intensity can thus be estimated to be of the order of $10^{19}$ erg s$^{-1}$cm$^{-2}$. FLUKA simulations show that such a positron beam co-propagates with an electron beam with an average density of about $n_e\approx2\times10^{15} $ cm$^{-3}$. The positron contribution on the leptonic beam will therefore be of the order of 10\% with a null component of positive ions. 

We compare now our experimental results with the electron-positron astrophysical jets. Even though a debate is still open as to whether these jets are predominantly constituted by an electron-proton plasma or by electron-positron pairs, an element in favor to the latter is the power-law continuum spectra of the gamma-ray bursts associated with these structures (indication of the jets being optically thin) \cite{Hoshino}. Despite the different generation mechanism (pair production from gamma-gamma instead of gamma-nucleus collisions), their composition would be similar to the jets reported here, also thanks to their co-propagation with a high-density gamma-ray beam of similar size and duration (FLUKA simulations indicate a gamma-ray brilliance of the order of $10^{19} - 10^{20} $ ph/s/mm$^2$/mrad$^2$/0.1\%BW). In the experimental results reported here, the excess of electrons in the beam implies a net current density of the order of $J_e \approx -(n_e^- -n_e^+) e c\approx -10^{11}$ A/m$^2$ (assuming $n_e\approx 2\times10^{15}$ cm$^{-3}$) inducing an azimuthal magnetic field of the order of $|B_{\phi}|\approx 30$ T. However, FLUKA simulations indicate that, by varying the electron beam characteristics and target thickness, it is possible to significantly modify this percentage. For instance, the interaction of a GeV-like laser-accelerated electron beam with a thicker tantalum target ($d \approx 2$ cm) is expected to generate a high-density, purely neutral electron-positron beam with an overall leptonic density of the order of $10^{16} - 10^{17}$ cm$^{-3}$. In this case, the toroidal magnetic fields would be virtually zero, allowing one to unveil the microphysics induced by the presence of small-scale magnetic fields generated by filamentation instability [11]. The high leptonic density would in fact allow for the study of the propagation of these jets in much more rarefied plasmas, in a scenario comparable to the propagation of these jets in the interstellar medium. 

The proposed mechanism for the generation of ultra-relativistic positron beams, applied to the near-term developments in laser technology, might also be relevant to the construction of all-optical electron-positron colliders. FLUKA simulations of the interaction of a pencil-like monoenergetic electron beam ($E_{e^-} =$ 100 GeV, overall charge of 1 nC, see \cite{Tzoufras})  with a 2 cm thick Ta target indicate the production of a positron beam with an exponentially decreasing energy spectrum (maximum energy of $E_{e^+} =$ 80 GeV, with approximately $10^6$ positrons with energy between 70 and 80 GeV), a divergence of the order of $\theta\approx10$ $\mu$rad and an overall charge comparable to that of the incoming electron beam. Particle-driven plasma wakefield accelerators can also be subsequently employed to further increase the positron energy in a metre-scale device, as recently demonstrated by Blumenfeld and collaborators \cite{Blumenfeld}. The extremely low-divergence achievable with our proposed generation mechanism would prove fundamental for efficient injection of the positrons into such devices. The normalised emittance of such a positron beam can be expressed, for each monoenergetic component, as $\varepsilon_n \approx \gamma\zeta\theta$, being $\zeta$ the beam source size. Due to the ultra-relativistic nature of the generation mechanism proposed, $\gamma\theta \approx 1$ rad regardless of the positron energy, thus reducing the normalised emittance to be $\varepsilon_n \approx \zeta \approx 30\pi$ mm mrad, in the conservative case of a $100$ $\mu$m source size. This is comparable to the positron emittance measured after the injection stage of LEP ($\varepsilon_{LEP} \approx 60\pi$ mm mrad \cite{LEPemittance}). It must also be noted that the positron beam would be inherently synchronised with the laser, allowing for the possibility of both electron-laser and positron-laser interactions. Direct comparison between these two experimental scenarios might allow for the testing of possible matter/anti-matter asymmetries in a highly non-linear regime.  
 
The authors acknowledge the funding schemes NSF CAREER (Grant 1054164) and NSF/DNDO (Award No. F021166). GS wishes to acknowledge the support from the Leverhulme Trust (Grant: ECF-2011-383). ADP is grateful to A. I. Milstein and to A. B. Voitkiv for stimulating discussions.

\end{document}